\documentclass[aps,pra,preprint,showpacs,amssymb,superscriptaddress,longbibliography,floatfix]{revtex4-1}

\usepackage{lineno,textcomp}

\usepackage{bm}
\usepackage{color}

\usepackage[german,english]{babel}

\usepackage{graphicx}
\graphicspath{{pict/}{figures/}{}}

\usepackage{amsmath,amssymb}

\usepackage[pdftex,colorlinks=true,bookmarks=false,citecolor=blue,urlcolor=blue]{hyperref} 
\hypersetup{pdfstartview={FitH},pdfpagemode={UseNone}, breaklinks=true,
            colorlinks,linkcolor=blue, citecolor=blue, urlcolor=blue,
             bookmarksopen=true, pdfnewwindow=true}

\newcommand\hrefBibPDF[3][]{}

\newcommand\reqtn[1]{\ref{eq:#1}}
\newcommand\reqt[1]{(\reqtn{#1})}

\newcommand {\grsim} {\ {\raise-.5ex\hbox{$\buildrel>\over\sim$}}\ }
\newcommand {\lessim} {\ {\raise-.5ex\hbox{$\buildrel<\over\sim$}}\ }

\newcommand{\invisiblesection}[1]{%
  \phantomsection%
  \stepcounter{section}%
  \addcontentsline{toc}{section}{\protect\numberline{\thesection}#1}%
  }

\begin{document}

\title{Extreme defect sensitivity from large synthetic dimensionality}

\author{Lukas J. Maczewsky}
\thanks{These authors contributed equally}
\affiliation{Institut f\"ur Physik, Universit\"at Rostock, Albert-Einstein-Straße 23, 18059 Rostock, Germany}

\author{Kai Wang}
\thanks{These authors contributed equally}
\affiliation{Nonlinear Physics Centre, Research School of Physics and Engineering, The Australian National University, Canberra ACT 2601, Australia}

\author{Alexander A. Dovgiy}
\affiliation{Nonlinear Physics Centre, Research School of Physics and Engineering, The Australian National University, Canberra ACT 2601, Australia}

\author{Andrey~E.~Miroshnichenko}
\affiliation{School of Engineering and Information Technology, University of New South Wales, Canberra, ACT 2600, Australia}

\author{Alexander Moroz}
\affiliation{Wave-scattering.com}

\author{Max Ehrhardt}
\author{Matthias Heinrich}
\affiliation{Institut f\"ur Physik, Universit\"at Rostock, Albert-Einstein-Straße 23, 18059 Rostock, Germany}

\author{Demetrios~N.~Christodoulides}
\affiliation{CREOL, University of Central Florida, 4304 Scorpius St., Orlando, Florida 32816 USA}

\author{Alexander Szameit}
\email{alexander.szameit@uni-rostock.de}
\affiliation{Institut f\"ur Physik, Universit\"at Rostock, Albert-Einstein-Straße 23, 18059 Rostock, Germany}

\author{Andrey A. Sukhorukov}
\email{andrey.sukhorukov@anu.edu.au}
\affiliation{Nonlinear Physics Centre, Research School of Physics and Engineering, The Australian National University, Canberra ACT 2601, Australia}






\date{\today}
\maketitle

\invisiblesection{Fully referenced paragraph - Nature style}
{\bf The geometric dimensionality of a physical system significantly impacts its fundamental characteristics~\cite{Ehrenfest:1920-440:RAR, Freeman:1969-1222:AMJP, Barrow:1983-337:PTRSA, Burgbacher:1999-625:JMP}. For example, the occurrence of Anderson localisation~\cite{Abrahams:1979-673:PRL, Schwartz:2007-52:NAT, Lahini:2008-13906:PRL, Conti:2008-794:NPHYS}, the stability of planetary orbits~\cite{Ehrenfest:1920-440:RAR, Barrow:1983-337:PTRSA}, and the number of components of the electric and the magnetic field~\cite{Ehrenfest:1920-440:RAR, Einstein:1938-683:ANNM} are a function of the number of the involved spatial dimensions. While experiments are fundamentally limited to the maximum of three spatial dimensions, there is a growing interest in harnessing additional synthetic dimensions. The numerous approaches to this end often involve the temporal and/or spectral domains and necessitate complex dynamic modulation regimes~\cite{Yuan:2018-1396:OPT}. Alternative methods relying on artificially engineered degrees of freedom generally do not scale well when more artificial dimensions are to be added. In our work, we introduce a new paradigm for the experimental realization of excitation dynamics associated with many-dimensional systems. Crucially, it relies solely on static one-dimensional equivalent structures with judiciously tailored parameters to faithfully reproduce the same optical spectrum and density of states of the high-dimensional system to be represented. In order to showcase the capabilities of our approach, we fabricate 1D photonic lattices that exhibit the characteristic non-monotonic excitation decays associated with quantum walks in up to 7D square lattices. Furthermore, we find that a new type of bound state at the edge of the continuum emerges in higher-than-three dimensions and gives rise to a sharp localisation transition at defect sites. In a series of experiments, we implement the mapped equivalent lattices of up to 5D systems and observe an extreme increase of sensitivity with respect to the detuning of the respective  anchor sites.
Our findings demonstrate the feasibility and applicative potential of
harnessing high-dimensional effects in planar photonics for ultra-sensitive switching or sensing.
Notably, our general approach is by no means limited to optics, and can readily be adapted to
a variety of other physical contexts, including cold atoms~\cite{Mancini:2015-1510:SCI, Stuhl:2015-1514:SCI} and superconducting qubits~\cite{Tsomokos:2010-52311:PRA} with exclusively nearest-neighbour interactions, promising to drive significant advances in different fields including quantum simulations and information processing.
}

\pagebreak
\invisiblesection{Introduction}
The concept of synthetic dimensions~\cite{Boada:2012-133001:PRL} is one of the current focal points of interest, since it effectively extends the scope of experimental observations beyond the (3+1)-dimensional space-time and thereby opens up new opportunities in various physical contexts, from innovative light control to quantum information processing~\cite{Yuan:2018-1396:OPT}.
Current approaches for synthesising dimensions commonly rely on either introducing additional parameter dependences~\cite{Zilberberg:2018-59:NAT} to the Hamiltonian, or using artificially engineered degrees of freedom, such as discrete frequencies, to form a synthetic space~\cite{Regensburger:2011-233902:PRL, Regensburger:2012-167:NAT, Mancini:2015-1510:SCI, Stuhl:2015-1514:SCI, Yuan:2016-741:OL, Yuan:2018-104105:PRB, Lustig:2019:NAT}, with latest experiments demonstrating up to four synthetic dimensions~\cite{Zilberberg:2018-59:NAT}.
Generally, the aim is to achieve a perfect one-to-one reproduction of the entire multi-dimensional lattice in synthetic space. However, this inevitably comes at the steep cost of exponentially increasing complexity when attempting to access higher effective dimensions.

Here we introduce a new paradigm for the selective realization of a wide range of useful physical effects associated with high-dimensional lattices and their nontrivial excitation dynamics. We show that arbitrary Hermitian multi-dimensional lattices with any local or non-local coupling distribution can be mapped to a 1D lattice with judiciously tailored nearest-neighbour couplings and detunings, such that the dynamics at a chosen anchor site is faithfully reproduced. Notably, in order to establish the effectively multi-dimensional environment in the 1D equivalent lattice, it is not sufficient to merely preserve the optical eigenvalue spectrum, as e.g. supersymmetry (SUSY) and related transformations~\cite{Miri:2014-89:OPT, Teimourpour:2016-33253:SRP, Yu:2016-836:OPT} are known to do. Instead, one has to exactly replicate the actual local density of states.
Since the method presented here does not require dynamic modulations, the static 1D equivalent lattices it yields are compatible with a broad range of existing technological platforms, including, but not limited to, the planar laser-written photonic circuits that we employ in our proof-of-concept experiments.

It is a well-known fact that excitation dynamics, in particular the emergence of defect-localised states, crucially depends on the lattice dimensionality, as dictated by the fundamental distinctions in density of states at the band-edges~\cite{Economou:2006:GreensFunctions}. As such, they are an ideal test case for any approach to synthetic dimensions. As it turns out, one of the key features of our technique is that any detuning-type defect at the anchor site identically impacts the excitation dynamics in the multi-dimensional structure and the synthetic 1D equivalent lattice. The approach presented in this work finally places these and many other high-dimensional phenomena within the experimentalist's grasp. In this vein, we observe the emergence of a sharp localisation transition at a critical defect strength. This behaviour is mediated by a new type of bound state at the edge of the continuum (BSEC) that occurs exclusively in four or more dimensions, which constitute a fundamentally nontrivial addition to the previously identified families of bound states in the continuum (BIC)~\cite{Hsu:2016-16048:NRM}. The resulting extreme defect sensitivity readily lends itself to multifarious practical applications and may serve as an entirely new design approach in metrology, e.g. for the precise measurement of minute optical refractive index contrasts.

\begin{figure}
    	\centering
 	\includegraphics[width=1\textwidth]{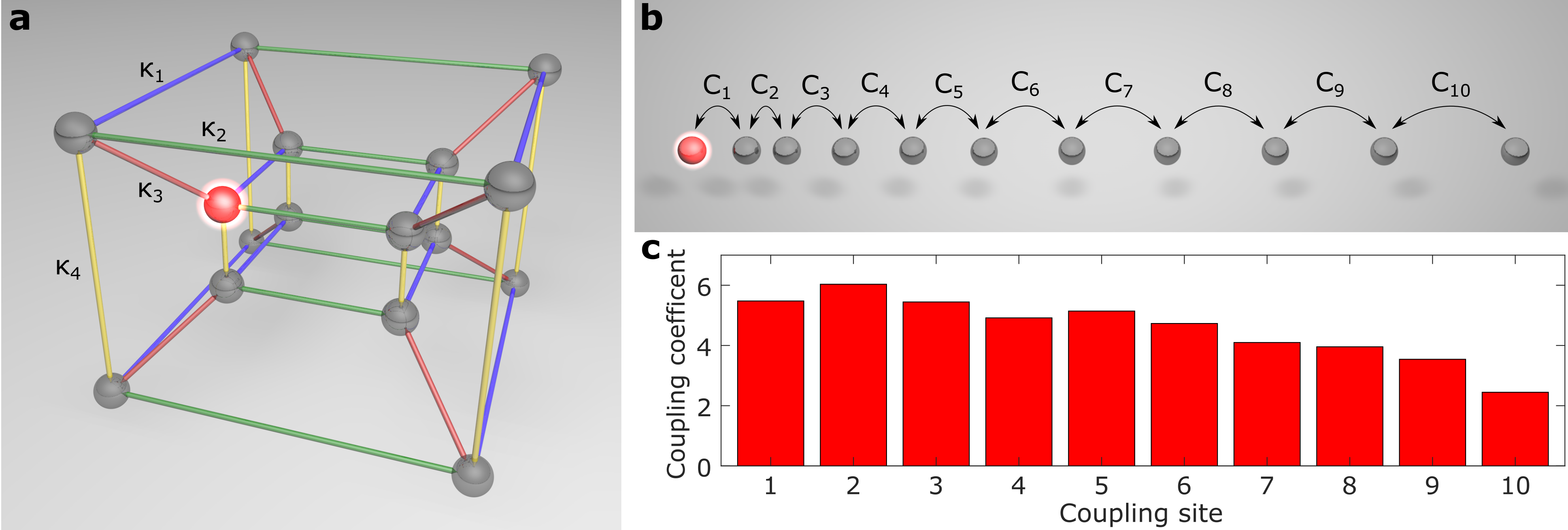}
 	\caption{\label{fig1} (a)~Sketch of a four-dimensional tesseract (3D projection of the structure), where spheres mark the discrete sites and the connecting lines between them 			indicate the couplings $\kappa_1=1,$ $\kappa_2=2,$ $\kappa_3=3,$ and $\kappa_4=4$. (b)~The 1D mapping of this 4D object, where the excitation dynamics of first site is the same as for the corresponding red anchor site in (a).
			(c)~The nearest-neighbour coupling coefficients $C_m$ of the one-dimensional mapped lattice. }
 \end{figure}

\invisiblesection{Mapping of a high-dimensional lattice to 1D}
Recent theoretical works employed Householder and SUSY-related transformations to map the optical spectrum of multi-dimensional arrangements to 1D lattices~\cite{Teimourpour:2016-33253:SRP, Yu:2016-836:OPT}. However, in those approaches, each site of the 1D lattice inevitably ends up representing a complex superposition of multiple original lattice sites. In contrast, we formulate a general mapping procedure to achieve a one-to-one correspondence between the respective anchor sites, which renders our approach intrinsically robust against arbitrary local defects. In order to illustrate this, we consider a lattice with arbitrary multi-dimensional geometry
governed by a linear Hermitian Hamiltonian $\textbf{H}$ and a local defect $\rho$ at site $m_\mathrm{d}$. Figure~\ref{fig1}(a) illustrates this for the example of a four-dimensional hypercube, or tesseract. The corresponding dynamics are therefore governed by a discrete Schr\"odinger equation,
\begin{align} \label{eq:Ndimension}
	\mathrm{i}\frac{\mathrm{d}\Psi_m}{\mathrm{d}t}
    = \sum_{m'} \textbf{H}_{m,m'} \Psi_{m'}\,
      + \rho \delta_{m,m_\mathrm{d}} \Psi_m ,
\end{align}
where $t$ is the normalised time coordinate, the index $m$ denotes the lattice sites, $\Psi_m$ are the complex wave function amplitudes, and $\delta_{m,m_\mathrm{d}}$ is a Kronecker delta function.
Using the Lanczos method~\cite{Lanczos:1950-255:RAR, Ojalvo:1970-1234:RAR}, we identify a unitary transformation $\textbf{V}$ that relates $\textbf{H}$ to a three-diagonal Hamiltonian $\textbf{H}^{(3)}=\textbf{V}^\ast \textbf{H} \textbf{V}$ corresponding to a 1D lattice with nearest-neighbour coupling. Considering single-site excitations at anchor site $m_\mathrm{d}$ as the initial step for a recursive Lanczos algorithm (see Supplementary for details),
the evolution Eq.~\reqt{Ndimension} can then be transformed to
\begin{align}
   \mathrm{i}\frac{\mathrm{d}\Phi_m}{\mathrm{d}t}=\epsilon_m \Phi_m + C_{m-1} \Phi_{m-1}+C_{m}\Phi_{m+1} + \rho \delta_{m,1} \Phi_m ,
\label{eq:coupledmode}
\end{align}
where for $m \ge 1$, $\epsilon_m$ and $C_m$ are the respective on- and off-diagonal elements of $\textbf{H}^{(3)}$, as illustrated in Figs.~\ref{fig1}(b,c).
It should be emphasised that the dynamics of an excitation at the anchor site in the multi-dimensional lattice (considering $\Psi_{m \ne m_\mathrm{d}}(t=0) = 0$) is directly mapped onto the dynamics of the first site in the 1D lattice (red sites in Figs.~\ref{fig1}(a,b)):
\begin{equation}
    \Phi_1(t) \equiv \Psi_{m_\mathrm{d}}(t) .
\end{equation}
Notably, this holds true for detunings $\rho$ of any strength, which can even be time-dependent and/or nonlinear according to an arbitrary function $\rho(t, \Psi_{m_\mathrm{d}})$ (see Supplementary for details).

\begin{figure}
 	\centering
 	\includegraphics[width=1\textwidth]{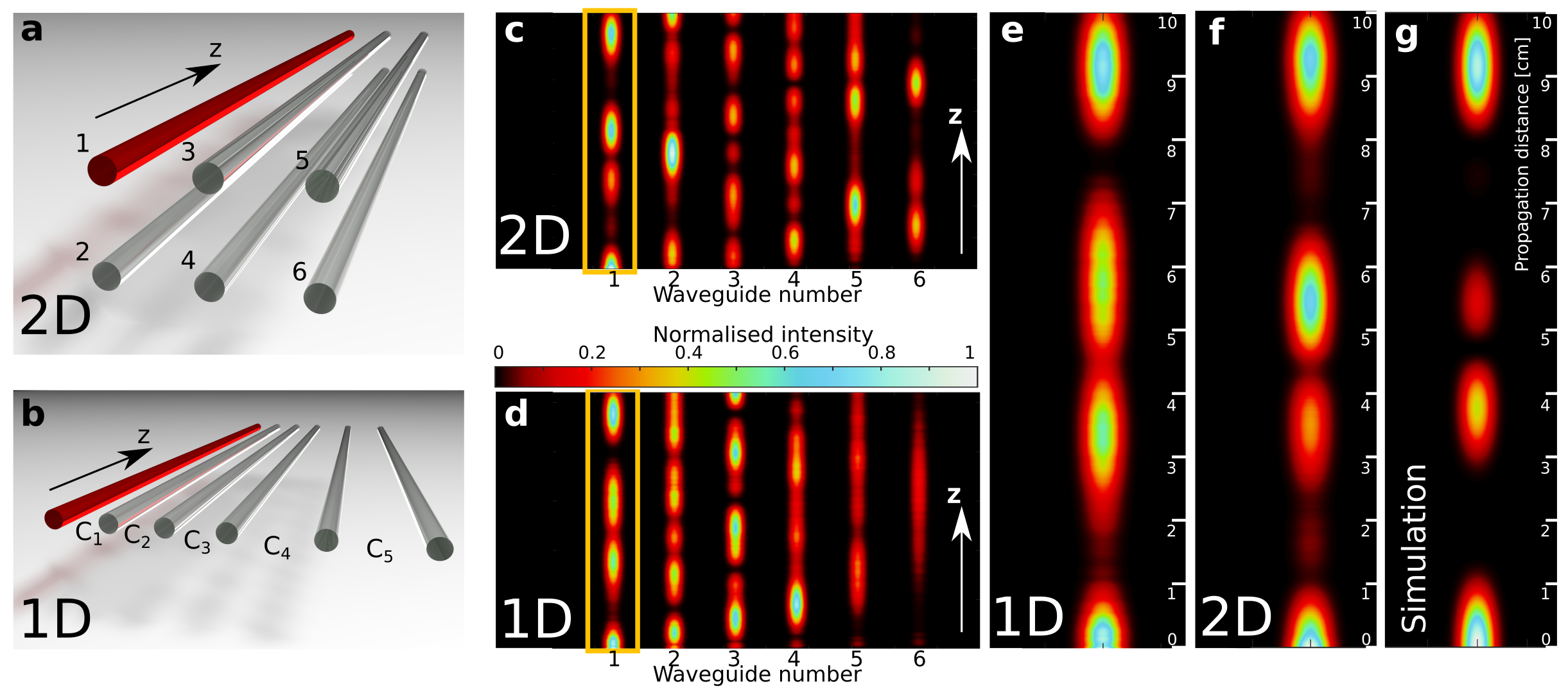}
 	\caption{(a) Schematic of a 2D square lattice with equal horizontal and vertical couplings, and (b) the mapped 1D lattice with its non-uniform coupling distribution $C_n$ 					implemented by different spacings.
 			(c,d)~Experimentally observed propagation dynamics in the experimentally realized 2D and 1D arrays. The arrows denote the propagation direction $z$ along the 						waveguides. The 2D structure has been inscribed at a tilt angle of $20^\circ$ to allow each lattice site to be viewed without obstruction or overlay.
 			(e) Zoomed-in views of the anchor waveguide in the 2D lattice, (f) the first waveguide in the 1D equivalent lattice, and (g) tight-binding simulation based on Eq.~(\ref{eq:coupledmode}) with the visualized transverse intensity distribution according to the waveguide mode profile.}
     \label{fig2}
 \end{figure}

We employed femtosecond laser-written photonic lattices~\cite{Szameit:2010-163001:JPB} as testbed for the experimental characterization of our equivalent structures.
Here, the propagation direction $z$ corresponds to the time coordinate $t$, and the strength of coupling between two adjacent waveguides can be continuously tuned by adjusting their transverse separation. In order to establish the viability of the mapping procedure, we directly compared the excitation dynamics in a 2D square lattice comprised of six waveguides (Fig.~\ref{fig2}(a)) with equal horizontal and vertical couplings and negligible diagonal interactions to the mapped 1D counterpart (Fig.~\ref{fig2}(b)). In the absence of detunings, the equivalent lattice features identical waveguides with nonuniform but constant couplings (see Supplementary).
Figures~\ref{fig2}(c,d) show the experimentally observed intensity evolution in the two systems obtained by fluorescence microscopy~\cite{Szameit:2010-163001:JPB}.
Remarkably, the dynamics of the anchor site (Fig.~\ref{fig2}(e)) are captured by the first site of the equivalent lattice (Fig.~\ref{fig2}(f)) so well that the observed intensity distributions in them are more similar to one another (fidelity $F=94.1\,\%$) than either of them is to the simulated behaviour based on the tight-binding approximation of Eq.~(\ref{eq:coupledmode}) (fidelities $F=79.2\,\%$ and $F=80.9\,\%$, respectively). This indicates that our method is not only robust with respect to detunings of the anchor site, but also can accommodate a certain amount of fabrication inaccuracies and systematic experimental deviations that have not explicitly been incorporated during the design process of the 1D equivalent lattice.

\begin{figure}
 	\centering
 	\includegraphics[width=1\textwidth]{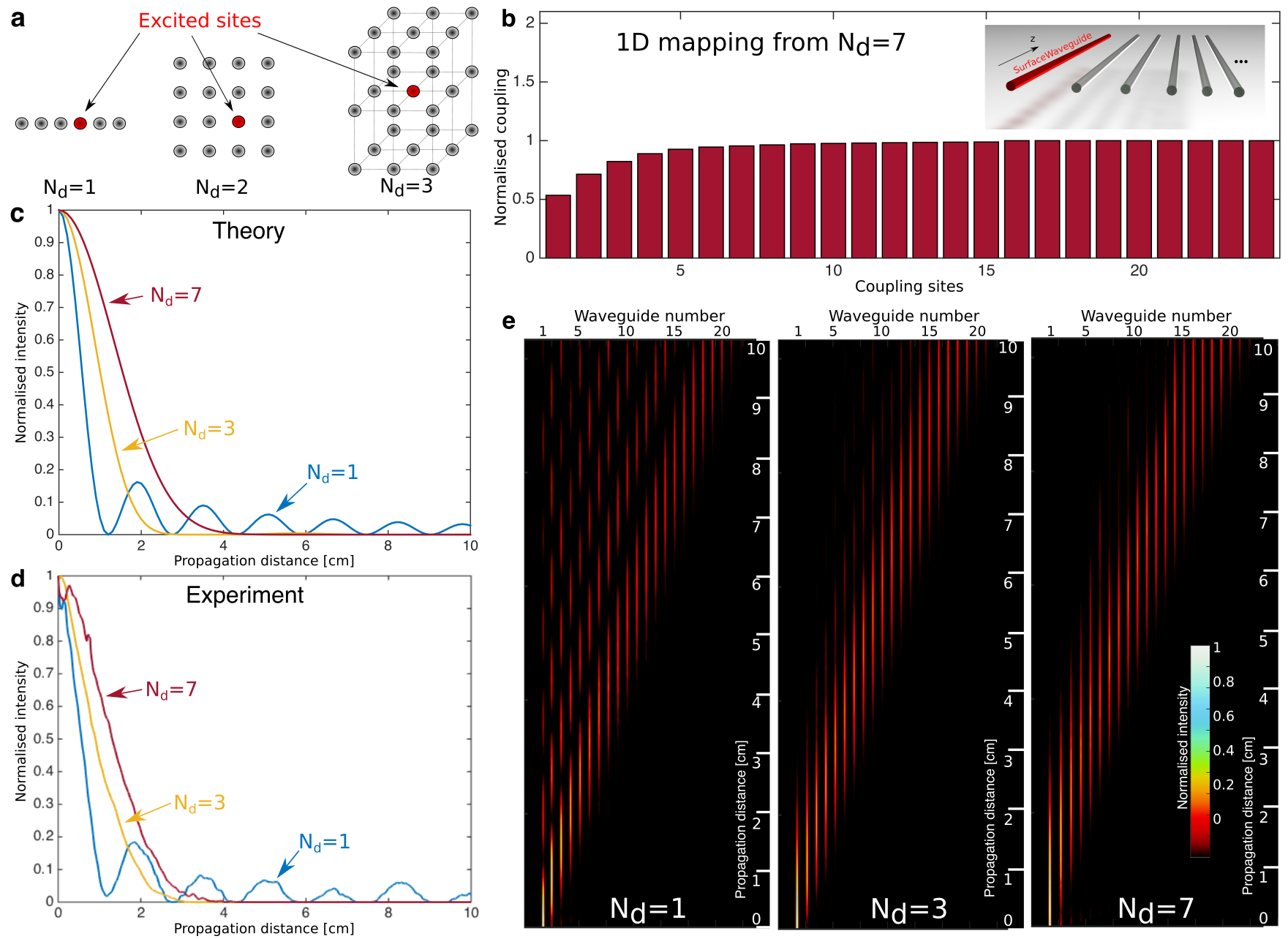}
 	\caption{(a)~Sketch of 1D, 2D and 3D lattices, each with an anchor site marked in red. (b)~Calculated normalised coupling coefficients ($C_m / \widetilde{C}$) at the first 24 sites of a one-dimensional structure mapped from an $N_\mathrm{d}=7$ dimensional lattice. Inset: Schematic of the experimental realisation with appropriate inhomogeneous spacings between adjacent waveguides. (c,d)~Dynamics of intensities at the excited sites for $N_\mathrm{d} = 1,3,7$: (c)~theory and (d)~experimental observations in the first waveguides of the mapped 1D lattices with $\widetilde{C}=1\; {\rm cm}^{-1}$. (e)~Fluorescence images of the propagation dynamics in the 1D mapped lattices corresponding to the dimensionalities $N_\mathrm{d} = 1, 3,$ and $7$.}
     \label{fig3}
 \end{figure}

\invisiblesection{High-dimensional quantum walk}
Having verified their viability, we harnessed our mapping technique to experimentally investigate dynamics associated with high dimensions.
We consider periodic lattices of increasing dimensionality $N_\mathrm{d}$ as illustrated in Fig.~\ref{fig3}(a). For $N_\mathrm{d}>3$, we analyse their higher dimensional generalisations with the nearest-neighbour couplings normalised as $\widetilde{C}_{N_\mathrm{d}} = \widetilde{C} / N_\mathrm{d}$. This ensures that the respective systems exhibit photonic bands in the same interval $-2 \widetilde{C} < \beta < 2 \widetilde{C}$ in order to allow for a direct comparison of effects associated with the multi-dimensional diffraction relation $\beta = 2 \widetilde{C}_{N_\mathrm{d}} \sum_{n=1}^{N_\mathrm{d}} \cos(k_n)$, where ${\bf k} = (k_1, k_2, \ldots, k_{N_\mathrm{d}})$ is the wave vector of the Bloch waves.
In calculating the parameters of the mapped 1D lattices, we find that $\epsilon_m \equiv 0$. The coupling of the 1D equivalent systems exhibit characteristic distributions for each dimension $N_\mathrm{d}$, and universally converge to $C_m/\widetilde{C} \rightarrow 1$ at $m \gg 1$ according to the width of the photonic band. As an example, Fig.~\ref{fig3}(b) depicts the distribution for  $N_\mathrm{d}=7$; additional cases for other dimensionalities are provided in the Supplementary.
After implementing the 1D equivalent lattices by fs laser direct inscription, we used them to shed light on unique features of higher-dimensional quantum walks by observing the respective escape dynamics of single-site excitations. Figures~\ref{fig3}(c) and~(d) compare theoretical predictions according to the expression $|\Phi_1(z)|^2 = [\mathcal{J}_0(2 \widetilde{C}_{N_\mathrm{d}} z)]^{2 N_\mathrm{d}}$ (see Supplementary) and experimental observations for three representative cases ($N_\mathrm{d}=1,3,7$), which coincide with an excellent fidelity of $F > 0.98\,\%$. The paradoxical impact of dimensionality on the quantum walk becomes apparent: Despite the increasing number of potential coupling destinations in higher-dimensional space, light initially tends to escape slower. Moreover, the observed propagation patterns in the 1D equivalent lattices show that the partial revivals known from 1D lattices are entirely suppressed in higher dimensions (Fig.~\ref{fig3}(e)).

\begin{figure}
 	\centering
 	\includegraphics[width=1\textwidth]{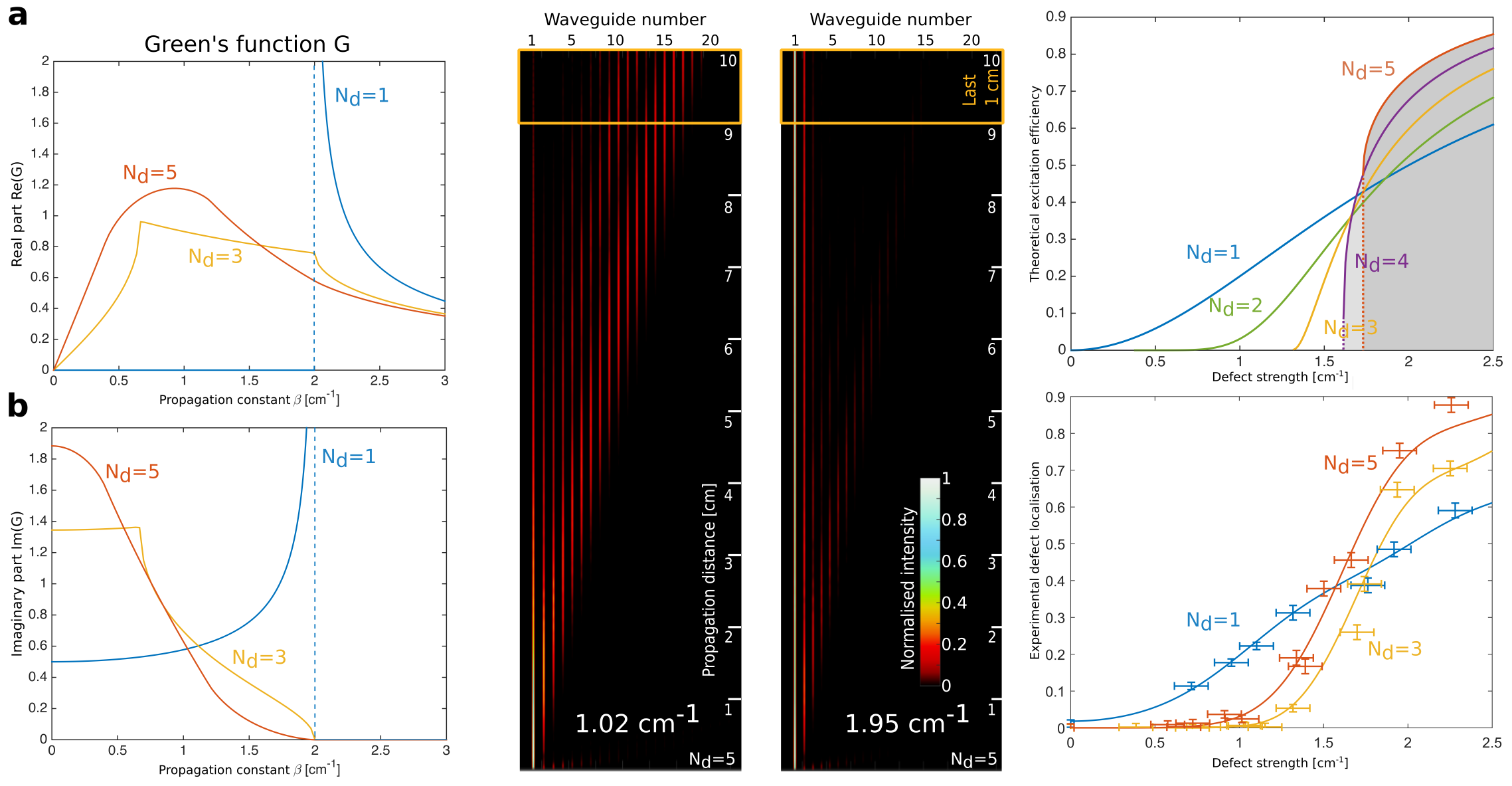}
 	\caption{(a),(b)~Real and imaginary parts of the spectral Green's function at the excitation site for lattices of different dimensionality $N_\mathrm{d} = 1,3,5$.
			(c),(d)~Fluorescence images of the mapped $N_\mathrm{d}=5$ dimensional lattices with a detuned anchor site: (c)~below the localisation threshold, for a detuning of $1.02\,					\mathrm{cm}^{-1}$, light is able to escape;
			(d)~above the localisation threshold, for a detuning  $1.95\,\mathrm{cm}^{-1}$, the initial excitation remains largely localized.
			(e)~Theoretically calculated excitation efficiency for infinite propagation distance. Shading outlines the region under the curves for $N_\mathrm{d} = 4,5$, where a distinct 				localization threshold occurs.
			(f)~Experimentally observed localisation at the defect (symbols) compared with the theoretical predictions (lines) in the planar lattices with different mapped 					dimensionality $N_\mathrm{d} = 1,3,5$, defined as the average relative intensity in the last $1\,\mathrm{cm}$ of the first waveguide. To indicate the influence of random 					fabrication imperfections, the error bars are numerically derived from an ensemble of 300 realizations with Gaussian-distributed lattice parameters centered around 				the design values with their typical standard deviation of $10\,\%$.}
     \label{fig4}
 \end{figure}

\invisiblesection{Highly-sensitive bound state at the continuum edge in higher dimensions}
From a mathematical point of view, the dynamics resulting from a single-site excitation represents the (spatial) Green's function $G_{N_\mathrm{d}}(z) = \Phi_1(z) / \Phi_1(z=0)$, which effectively characterises the response of the entire lattice. Its Fourier partner, the spectral Green's function,
\begin{equation}
    \widetilde{G}_{N_\mathrm{d}}(\beta) = -\mathrm{i} \int_0^{+\infty} G_{N_\mathrm{d}}(z)\ \mathrm{e}^{\mathrm{i} \beta z} \ \mathrm{d}z,
\end{equation}
provides additional insights into the role of dimensionality. Figures~\ref{fig4}(a,b)
show the theoretically calculated real and imaginary parts of $\widetilde{G}_{N_\mathrm{d}}$ for $N_\mathrm{d}=1,3$ and $7$. The imaginary part ${\rm Im}(\widetilde{G}_{N_\mathrm{d}}(\beta))$ determines the local density of states (LDOS) and is therefore responsible for the rate of escape from the excited site mediated by the different spectral components within the photonic band. The evident lower rate of escape observed in the mapped versions of higher-dimensional lattices is consistent with the calculated behaviour of the Green's function within the transmission band ($|\beta| < 2 \mathrm{cm^{-1}}$, up to the dashed lines in Fig.~\ref{fig4}(a,b)). In this vein, our mapping approach can also serve as versatile tool for engineering the band structure and tailoring the associated lattice response towards practical applications.

We finally consider the fundamental effect of localisation arising from a detuning $\rho$ of the anchor with respect to the surrounding homogeneous $N_\mathrm{d}$-dimensional lattice.
Importantly, knowledge of the Green's function in a defect-free lattice allows us to predict the localisation behaviour upon introduction of a defect~\cite{Economou:2006:GreensFunctions} as
\begin{equation}
    \rho = 1 / \widetilde{G}_{N_\mathrm{d}}(\beta_\mathrm{d}) ,
\end{equation}
where $\beta_\mathrm{d}$ is the propagation constant of a localised defect mode.  It follows that real-valued detunings $\rho$ can support localization in the photonic band-gap, where ${\rm Im}\,\widetilde{G}_{N_\mathrm{d}}(\beta) = 0$. The real part of the Green's function accordingly reflects this tendency of light to remain localised in the excited anchor site. We determine the excitation efficiency for the defect-localised mode, defined as the fraction of power remaining at the input site after a sufficiently long evolution. In the theoretical limit of $z \rightarrow +\infty$, it is expressed as (see Supplementary):
\begin{equation}\label{eq:ee}
   \eta= \left(\frac{\partial \rho}{\partial \beta_\mathrm{d}}\right)^{-2}.
\end{equation}
In Fig.~\ref{fig4}(e) we show the calculated excitation efficiency over the defect strength for the dimensionalities from 1 to 5.
In 1D and 2D, localised modes appear for any defect strength, while for $N_\mathrm{d} \ge 3$ localisation occurs only above a critical defect magnitude, $\rho \ge \rho_{\rm cr}$, in agreement with previous studies~\cite{Economou:2006:GreensFunctions}.
Remarkably, we identify a distinctive effect for high-dimensional lattices with $N_\mathrm{d} \ge 4$: Here, the excitation efficiency instantly attains a finite non-zero value at the critical defect strength. Specifically, we see in Fig.~\ref{fig4}(e) that  $\eta(\rho_{\rm cr}) \simeq 0.13$ for 4D and $\simeq 0.49$ for 5D lattices.
The key aspect in allowing this characteristic of a high-dimensional system to be mapped to a 1D structure is that the judiciously synthesized coupling distribution of the latter likewise supports a bound state at the edge of the continuum (BSEC) exactly for the critical defect strength (see Supplementary).
Since the associated sharp transition from zero to strong localisation had not been observed before, we experimentally explored the phenomenon  in the respective 1D equivalent lattices, where the anchor site's detuning was implemented by choosing a different inscription speed (see Supplementary).
The excitation efficiencies were estimated by averaging the final centimetre of the recorded fluorescence images (Fig.~\ref{fig4}(c,d)), as indicated by orange frames, and calculating the fraction of light remaining in the anchor site. As an example, for $N_\mathrm{d}=5$ the light remains much more strongly trapped for an above-threshold detuning (Fig.~\ref{fig4}(c,d)). Figure~\ref{fig4}(f) shows an overview of the localization transition for different mapped dimensionalities. Note that, while the finite sample length of $10\,\mathrm{cm}$ in our experiments necessarily smooths out the abrupt jump predicted for infinite propagation distance, the systematically increased slope clearly showcases the dramatically enhanced sensitivity in higher dimensions. Mapped equivalent lattices provides the means to harness this extreme sensitivity, as well as many other unique features of higher-dimensional systems, in a single site of an appropriately engineered 1D system.

\invisiblesection{Discussion}
In summary, we propose and demonstrate a general approach for mapping the physics of arbitrary high-dimensional lattices onto exclusively one-dimensional equivalent lattices with nearest-neighbour interactions. By judiciously engineering the coupling and on-site energy profile, our approach goes beyond globally matching the eigenenergies of the multi-dimensional system, and faithfully reproduces the actual local density of states at the anchor site. This innovative concept not only allows us to observe the dynamical features of quantum walks in record high lattice dimensions, but also opens up new opportunities to harness otherwise inaccessible high-dimensional features for practical applications. As an important usage case, we demonstrate an extreme increase of sensitivity associated with the sharp localization transition at surface defects in four- or higher-dimensional lattices, facilitated by a new type of bound state at the edge of continuum in a mapped 1D lattice, and propose its application in enhanced metrology designs.
The platform-independent nature of our approach will boost the development of tailored one-dimensional lattices in a variety of systems and enable previously inaccessible exotic high-dimensional effects to find their way to applications in many different fields such as topological photonics, quantum simulation and optical information processing. In this vein, 1D mapped equivalent lattices hold great promise for the exploration of high-dimensional physics under linear, nonlinear and even non-classical conditions. Long-standing questions such as the dimensional scaling behavior of Anderson-localization, the stability of solitons, and entanglement dynamics in higher dimensions are now in reach of tabletop experiments.

\section*{Acknowledgements}
This work was supported by the Australian Research Council (DP160100619); the Australia-Germany Joint Research Cooperation Scheme; Erasmus Mundus (NANOPHI 2013 5659/002-001); the Alexander von Humboldt-Stiftung, and the German Research Foundation (BL 574/13-1, SZ 276/15-1). A.S. gratefully acknowledges financial support from the Alfried Krupp von Bohlen und Halbach Foundation. K.W. acknowledges fruitful discussions with Shanhui Fan and financial support from SPIE Optics and Photonics Education Scholarship. The authors would also like to thank C.~Otto for preparing the high-quality fused silica samples used in all experiments presented here.

\subsection*{Author contributions}
The theory was developed by Kai Wang, Alexander Dovgiy, Andrey Miroshnichenko, Alexander Moroz, Demetrios Christodoulides, and Andrey Sukhorukov. The design, implementation and characterisation of the lattice structure was done by Lukas Maczewsky, Max Ehrhardt, Matthias Heinrich, and Alexander Szameit. The project was supervised by Alexander Szameit and Andrey Sukhorukov. All authors discussed the results and co-wrote the paper.

\subsection*{Competing interests}
The authors declare no competing interests.

\section*{Methods}

\subsection{Sample fabrication and  fluorescence microscopy}
The photonic lattices for our experiments were manufactured by the femtosecond laser writing method~\cite{Szameit:2010-163001:JPB}. The Titanium:sapphire amplifier system (\textsc{Coherent} Mira/RegA) supplied $150\,\mathrm{fs}$ pulses with an energy of $250\,\mathrm{nJ}$ and a carrier wavelength of  $800\,\mathrm{nm}$ at a repetition rate of $100\,\mathrm{kHz}$. A microscope objective ($0.35\,\mathrm{NA}$) was employed to focus these pulses int the bulk of a $100\,\mathrm{mm}$ long fused silica sample (\textsc{Corning} 7980), which in turn was translated at speeds of approximately $100\,\mathrm{mm}\,\mathrm{min}^{-1}$ by a high-precision positioning system (\textsc{Aerotech} ALS 130). With an index contrast on the order of $7\times 10^{-4}$, the resulting waveguides exhibit a mode field diameter of $10.4\,\text{\textmu m} \times 8\,\text{\textmu m}$ at $633\,\mathrm{nm}$. Propagation losses and birefringence were estimated at $0.2\,\mathrm{dB}\,\mathrm{cm}^{-1}$ and $10^{-7}$, respectively. Details on the dependence of coupling and waveguide detuning on the fabrication parameters are given in the Supplementary.

Notably, the color centers formed during the inscription process enable the direct obervation of the propagation dynamics by means of fluorescence microscopy~\cite{Szameit:2010-163001:JPB}. Despite their low concentration and the resulting small conversion rate, meaningful quantitative measurements are enabled by separating stray excitation light ($633\,\mathrm{nm}$ Helium-Neon laser) from the fluorescence signal ($650\,\mathrm{nm}$). Typically, this technique has been limited to planar arrangements. To circumvent this limitation, we inscribed the two dimensional lattices in our experiments at a $20^\circ$ tilt, such that no two waveguides overlap vertically and the propagation dynamics of all lattice sites can be simultaneously imaged in a distinct fashion.

%

\end{document}